\documentclass[letter,longauth]{aa}  
\usepackage[utf8]{inputenc}
\usepackage{threeparttable} % to use table notes
\usepackage{graphicx}
\usepackage{natbib}
\usepackage{lipsum}
\usepackage{refcount}
\usepackage{txfonts}
\usepackage[colorlinks=true, allcolors=blue]{hyperref}

\begin{document}

   \title{Observation of an accreting planetary-mass companion with signs of disc–disc interaction in Orion~\thanks{Based on observations made with ESO Telescopes at the La Silla Paranal Observatory under programme ID 114.28JY}}
   \titlerunning{Planetary-mass companion in Orion}

   \author{ E. Vila\inst{1}
          \and P. Amiot\inst{1}
          \and O. Berné\inst{1}
          \and I. Schroetter\inst{1}
          \and T. J. Haworth\inst{2} 
          \and P. Zeidler\inst{3}
          \and C. Boersma\inst{4}
          \and J. Cami\inst{5, 6, 7}
          \and A. Fuente\inst{8}
          \and J.R. Goicoechea\inst{9}
          \and T. Onaka\inst{10}
          \and E. Peeters\inst{5, 6, 7}
          \and M. Robberto\inst{11, 12} 
          \and M. Röllig\inst{13, 14}
          }
    \authorrunning{E. Vila, P. Amiot, O. Berné, I. Schroetter, T. Haworth, P. Zeidler}

   \institute{Institut de Recherche en Astrophysique et Planétologie, Université de Toulouse, Centre National de la Recherche Scientifique (CNRS), Centre National d’Etudes Spatiales, 31028 Toulouse, France \\ \email{paul.amiot@gmail.com}  \and
   Astronomy Unit, School of Physics and Astronomy, Queen Mary University of London, London E1 4NS, UK  \and
   AURA for the European Space Agency, ESA Office, STScI, 3700 San Martin Drive, Baltimore, MD 21218, USA \and
   NASA Ames Research Center, MS 245-6, Moffett Field, CA 94035-
   1000, USA \and
   Department of Physics and Astronomy, University of Western Ontario, London, Ontario N6G 2V4, Canada
   \and 
   Institute for Earth and Space Exploration, University of Western Ontario, London, Ontario N6A 5B7, Canada
   \and 
   SETI Institute, Mountain View, CA 94043, USA
   \and
   Centro de Astrobiología (CAB), INTA-CSIC, Carretera de Ajalvir Km. 4, Torrejón de Ardoz, 28850 Madrid, Spain
   \and 
   Instituto de Física Fundamental (Consejo Superior de Investigacion
    Cientifica), 28006, Madrid, Spain
    \and
    Department of Astronomy, Graduate School of Science, The University of Tokyo, 7-3-1 Bunkyo-ku, Tokyo 113-0033, Japan 
   \and
   Space Telescope Science Institute, 3700 San Martin Dr., Baltimore, MD 21218, USA \and
   Johns Hopkins University, Bloomberg Center for Physics and Astronomy, 3400 N. Charles Street, Baltimore, MD
   21218, USA \and
   Physikalischer Verein, Gesellschaft f\"ur Bildung und Wissenschaft, Robert-Mayer-Str. 2, 60325 Frankfurt am Main, Germany \and
   Institut für Angewandte Physik, Goethe-Universität Frankfurt, Frankfurt am Main, Germany
             % \thanks{Shows the usage of elements in the author field}
             \\ }
\date{Received 8 September 2025; accepted 27 October 2025}

\abstract
{Young ($\lesssim 10$ Myr) planetary-mass companions (PMCs) provide valuable insights into the formation and early evolution of planetary systems. To date, only a dozen such objects have been identified through direct imaging. Using JWST/NIRCam observations towards the Orion Nebula, obtained as part of the \textit{PDRs4All} Early Release Science program, we have identified a faint point source near the M-type star V2376 Ori, a member of Orion D, around 80\,pc in the foreground of the Trapezium cluster of Orion and with an age of approximately $7 \pm 3$ Myr. Follow-up spectroscopic observations with the MUSE instrument on the VLT confirm that the source, V2376 Ori b, is indeed a young PMC.  We fit the SED of V2376 Ori b and infer a mass of $ \sim 20~M_{\rm Jup}$ {with evolutionary tracks}. The MUSE spectrum reveals several accretion tracers. Based on the H$\alpha$ line intensity, we estimate an accretion rate of $\sim$10$^{-6.6 \pm 0.2}~\rm M_{Jup}\,yr^{-1}$, which is comparable to that of young PMCs such as PDS~70b. In addition, the MUSE data cube reveals extended emission in the [O\,\textsc{ii}] doublet at 7320 and 7330~\AA, which is interpreted as evidence of a dynamical interaction between the two sources that potentially involves mass transfer between their individual accretion discs. These results demonstrate that JWST/NIRCam imaging surveys of young stellar associations can uncover new PMCs, which can then be confirmed and characterized through ground-based spectroscopic follow-up.
}

\maketitle

%-------------------------------------------------------------------

\section{Introduction}

Direct imaging has revealed key properties of planetary-mass companions (PMCs, $5 \lesssim M_p \lesssim 25~\rm{M}_{Jup}$, $ 10 \lesssim a \lesssim 300$~au; \citealt{currie_images_2022}). Only a dozen have been identified in young ($<10$ Myr) stellar clusters \citep[e.g.][]{Currie_2013, Luhman_2006, schmidt2008direct, Bowler_2014}. Despite their rarity, young PMCs offer valuable constraints on planetary formation pathways \citep{Kratter_2016, Schlaufman_2018}, especially in environments where external factors such as photoevaporation and dynamical interactions may play a role \citep{reiter_parker_2022, collaboration_past_2025}.
Expanding the sample of known young PMCs is critical but has been limited by the time-intensive nature of ground-based high-contrast imaging. JWST has recently demonstrated its potential to image both known and new PMCs as well as exoplanets \citep{Boccaletti_2024_HR8799, Matthews_2024_EpsilonIndiAb, carter_jwst_2023, Mullally_2024, lagrange_evidence_2025}.
In this letter, we report the discovery of a $\sim$7 Myr-old PMC in Orion, identified in JWST/NIRCam data from the \textit{PDRs4All} Early Release Science program \citep[PID 1288; ][]{berne_pdrs4all_2022, habart_pdrs4all_2024}, and confirmed via follow-up spectroscopy with VLT/MUSE.

\begin{figure*}
    \centering
    \includegraphics[width=0.8 \linewidth]{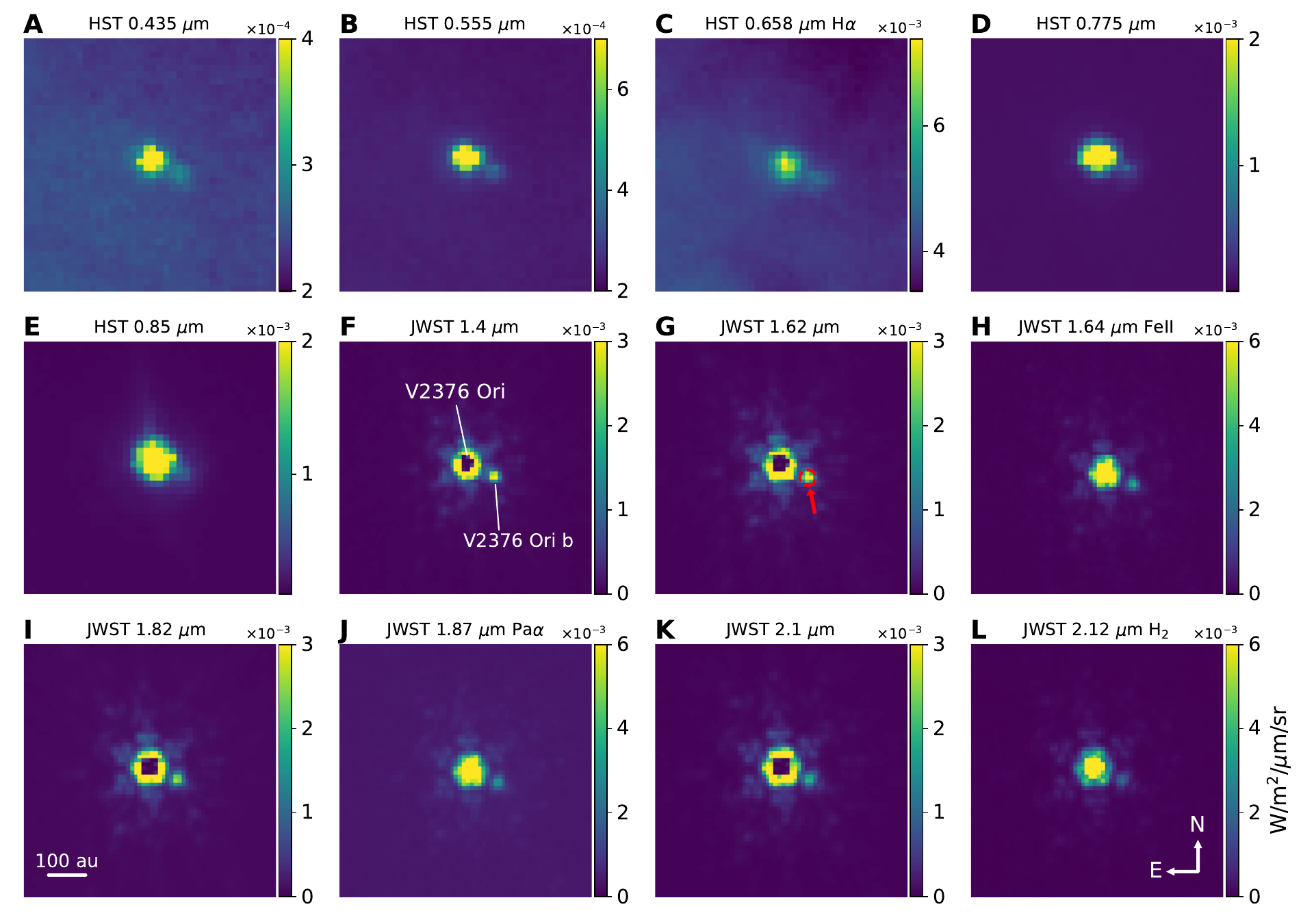}
    \caption{
    V2376 Ori observations.
    (\textbf{A} to \textbf{E}) HST ACS/WFC optical image \citep{ricci_hubble_2008}. (\textbf{F} to \textbf{L}) JWST near-infrared images \citep{habart_pdrs4all_2024}. 
    % The white arrow in panel (\textbf{G}) points to a red-dashed circle that indicates the aperture shape used to extract the V2376 Ori b fluxes. 
    The dashed red circle (pointed by the red arrow) in panel (\textbf{G}) indicates the aperture shape used to extract the V2376 Ori b fluxes. 
    The wavelength and physical assignment of each image are labelled above each panel for the narrowband filters. Black pixels are saturated. 
    }
    \label{fig:kaleidoscope}

\end{figure*}

\section{Source and data}

V2376 Ori ($\rm ra =83.83981804475 $, $\rm dec = -5.35155088077$) is a young star situated north-east of the Orion nebula cluster (ONC), catalogued as an M-type star by \citet{hillenbrand1997}.
In their catalogue of protoplanetary discs, \citet{ricci_hubble_2008} identified a faint source within $1''$ of V2376 Ori, which the authors interpreted as a jet.
This faint source was also reported by \citet{habart_pdrs4all_2024} in the Paschen $\alpha$ image obtained with the JWST. 
In these earlier studies, V2376 Ori was believed to be part of the ONC association. However, Gaia observations provide a parallax for V2376 Ori of $\pi = 3.29 \pm 0.46$ mas, placing the source at a distance of $d=304 \pm 42$ pc, which is much closer than the ONC (390 pc, \citealp{maiz_apellaniz_villafranca_2022}). This implies that V2376 Ori is instead part of the Orion D association identified by \citet{kounkel_apogee-2_2018}, a population of young stars (7-10 Myr) situated in front of the ONC. This foreground population has also been identified by \citet{Alves_2012_age} as a region dominated by the NGC 1980 cluster, aged 4-5 Myr. Thus, we assume an age of $7 \pm 3$ Myr for V2376 Ori.

We use JWST NIRCam images of the ONC, \textit{Hubble} Space Telescope (HST) images of V2376 Ori and new observations from VLT MUSE. The data presentation and reduction are presented in Appendix \ref{sect:data}.

\section{Results}

\begin{figure*}
    \centering
    \includegraphics[width=0.85\linewidth]{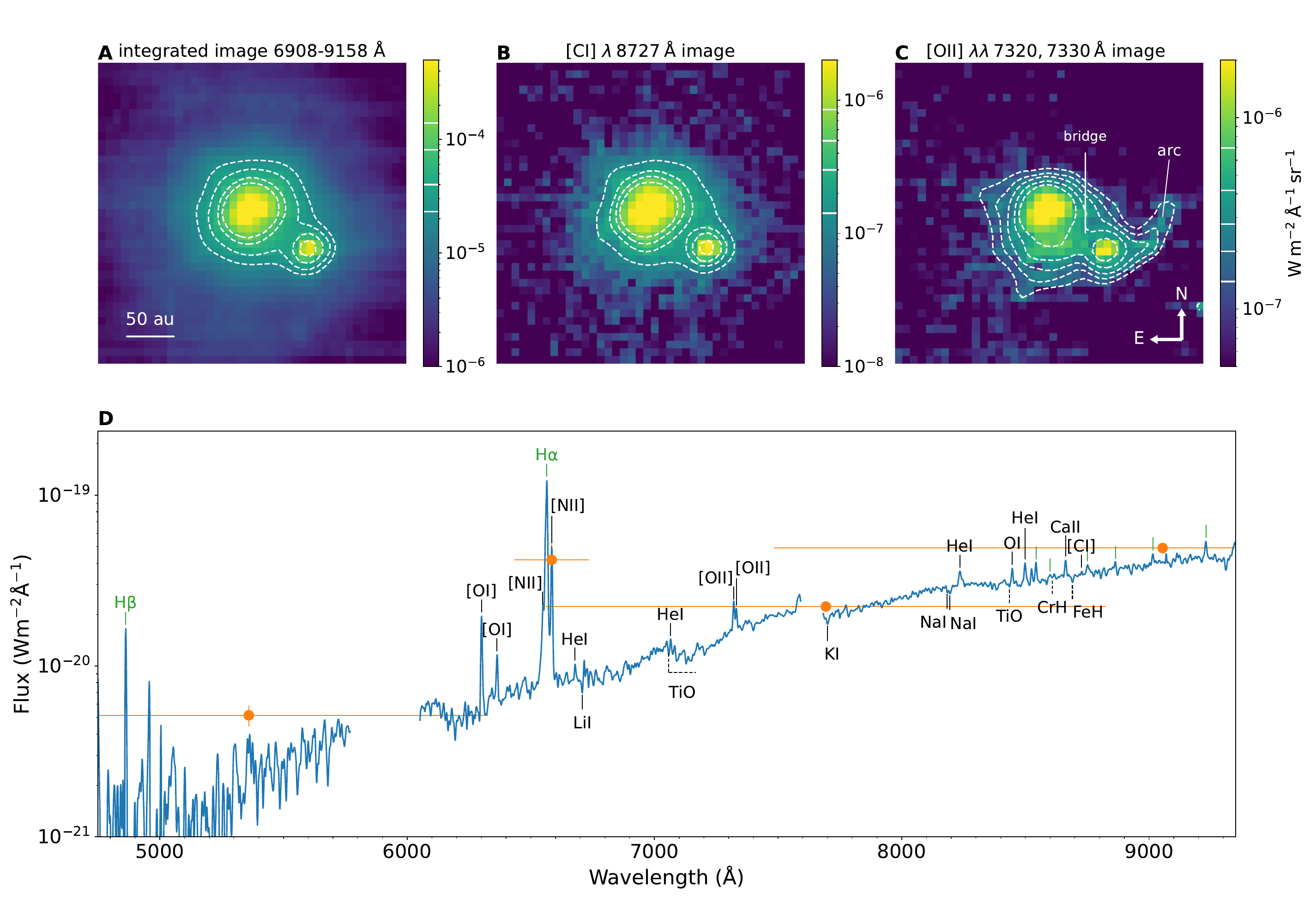}
    \caption{(\textbf{A} to \textbf{C}) Continuum-subtracted maps. (\textbf{A}) Integrated cube (6908-9158~\AA). (\textbf{B}) [\ion{C}{i}]~$\lambda$8272 integrated map. (\textbf{C}) Sum of the integrated maps of the [\ion{O}{ii}] doublet at 7320 and 7330~\AA.
    Dashed white contours indicate intensity levels. Their spacings are indicated with the whites lines on the colour bars.
     (\textbf{D}) MUSE spectrum (blue) and HST photometry (orange dots) of V2376 Ori b. HST filter widths are indicated by orange bars. Green lines mark recombination features; other lines are labelled in black.
    }
    \vspace{-0.5cm}
    \label{fig:muse}
    
\end{figure*}

Figure \ref{fig:kaleidoscope} presents the HST and JWST images of V2376 Ori obtained in 12 filters. The faint source previously reported by \citet{ricci_hubble_2008} and \citet{habart_pdrs4all_2024} is clearly detected in all bands.
Figure ~\ref{fig:muse} shows the MUSE integral field spectroscopic observations of the system. It combines integrated maps and the spectrum of the companion over the full MUSE wavelength range. Panel~\ref{fig:muse}{\bf A} displays the white light image integrated over the 6908–9158~\AA{} range. The companion source is clearly visible. Panel~\ref{fig:muse}{\bf B} shows the continuum-subtracted map (using the Background2D procedure described in Appendix \ref{sect:PSF_sub}) of the forbidden [\ion{C}{i}]~$\lambda$8727 line from neutral atomic carbon, where the companion is also detected. In addition, we find extended emission in the [\ion{O}{ii}] doublet at 7320 and 7330~\AA{}, which is shown in panel~\ref{fig:muse}{\bf C} as the sum of the integrated maps of both lines. A similar, although fainter, structure is also visible in the [\ion{O}{i}]~$\lambda$6300 map.
We analyse the relative astrometry of V2376 Ori and the nearby point source in Appendix~\ref{sect:rel_position}. We measure a angular separation of $0.23 \pm 0.03''$, corresponding to a projected separation of $a =71 \pm 21$~au when assuming a distance of $304 \pm 42$~pc. The relative position of the two objects has remained constant within instrumental uncertainties over $\sim$20 years, consistent with the source being gravitationally bound. We therefore identify it as a companion, hereafter referred to as V2376 Ori~b.
Panel ~\ref{fig:muse}{\bf D} presents the MUSE spectrum of this companion.

\subsection{Temperature, luminosity, and mass of V2376 Ori b}\label{sect:temperature_lumin_mass}

Figure~\ref{fig:SED} presents the HST and JWST spectral energy distribution (SED) of V2376 Ori b, obtained as explained in Appendix \ref{sect:PSF_sub}, within the dashed red circular aperture of 0.08'' radius, shown in Fig.~\ref{fig:kaleidoscope} ({\bf G}).
We fitted a black-body using a grid of temperatures ranging from 1000 K to 3000 K with a step of 100 K,  and a scaling coefficient, with values ranging from $10^{-27}$ to $10^{-20}$. 
The best-fit parameters are those that minimize the root mean square error (RMSE).
The error bars were determined considering the parameters range for which the RMSE is within 1.5 times the minimum RMSE. 
This yielded $T_{\rm eff} = 2100^{+200}_{-300}$ K.
The derived black-body luminosity is $ L = 10^{-2.4\pm 0.1}~ \text{L}_{\odot}$, where the uncertainty arises from the errors on both the temperature and the distance to the source. 
{As the system is situated in a young association, extinction from interstellar or circumstellar dust may affect the SED fitting. To investigate this, we used the extinction curve of \citet{Weingartner_Draine_2001} with $R_\text{V} = 5.5$ parametrized by the visual extinction, $A_{\rm V}$, in the fit. The best fit gives $A_{\rm V}$=0 suggesting a negligible effect of extinction.   
In Fig.~\ref{fig:SED} we also test the use of the BT-Cond model \citep{Allard_2001} to fit the SED; however, this gives poorer fits to the SED (see also the detailed comparison in Appendix~\ref{sect:models})}.
Figure~\ref{fig:evolutionary_tracks} shows the position of V2376 Ori b on the evolutionary tracks from \citet{baraffe_evolutionary_2003}. Considering the obtained luminosity, effective temperature, and an age of 7$\pm$3 Myr, this yields a mass in the range of 10-30 $\rm M_{Jup}$.
The derived physical properties of V2376 Ori b are summarized in Table~\ref{tab:Companion_prop}.
Figure~\ref{fig:exoplanets} shows that these properties are similar to those found for other directly imaged PMCs and giant exoplanets.

\subsection{MUSE spectroscopy of the V2376 Ori system}

Figure \ref{fig:manjavacas} presents the MUSE spectrum of V2376 Ori (see details on the extraction method in Appendix \ref{sect:linelist}). This spectrum is found to be compatible with that of an M7.5-9 spectral type (Appendix \ref{sect:host_star}).
Figure~\ref{fig:muse} (\textbf{D}) presents the MUSE spectrum for V2376 Ori b.
It shows smooth continuum emission, with an increase towards the red and numerous emission lines. This spectrum is similar to that of the MUSE spectrum of the $\sim$12 M$_{\rm Jup}$ PMC Delorme 1 (AB; \citealt{eriksson_strong_2020}). It shows broad, but weak, molecular absorption features in comparison to what is seen in isolated planetary-mass objects \citep[e.g.][]{manjavacas_spectral_2020, kirkpatrick_discovery_2006}. In Appendix \ref{sect:models} we fit this spectrum using atmospheric models and show that no model performs better than a simple black-body, highlighting the fact that models are not yet adapted for these young objects.
One striking aspect of the spectrum is the large number of emission lines (also not present in models), whose properties are summarized in Table~\ref{tab:linelist}.
These lines are commonly seen in young stellar objects and discs, and some of them have also been reported in PMCs (see details in Appendix \ref{sect:linelist}).
Combined with the young age of the system ($7\pm3$ Myr), these features strongly suggest that accretion is ongoing in V2376 Ori b. Interestingly, in the SED presented in Fig.~\ref{fig:SED}, the filter F187N (Paschen $\alpha$) shows an excess with respect to the reference broad filter F182M, suggesting that Pa$\alpha$ emission, another accretion tracer, is also present. From the H$\alpha$ and Pa$\alpha$ emission, we estimate an accretion rate $\dot{M}_{\rm acc}\simeq10^{-6.5\pm0.7}\:\rm M_{Jup}/yr$ (see Appendix \ref{sect:accretion}). This value is comparable to those found for other PMCs \citep{eriksson_strong_2020, Zhou_2014, Bowler_2011, Miranda_2018, Hashimoto_2020} and PMOs \citep{Mohanty_2005, Fang_2009, joergens_ots44_2013} with similar masses.

\subsection{Extended structures}

Figure~\ref{fig:muse} \textbf{C} shows the sum of the integrated maps of the [\ion{O}{ii}] doublet at 7320 and 7330~\AA. These auroral lines indicate tenuous, ionized gas at a few thousand Kelvin. Considering that V2376 Ori is $\sim 90$ pc in front of the ONC, it seems unlikely that the gas is ionized by extreme UV  ($h\nu>13.6$ eV) radiation from the ONC massive stars. Instead, it probably reflects ionization by local sources of extreme UV (e.g. from the accretion shocks). The extended [\ion{O}{ii}] emission (Fig.~\ref{fig:muse}) shows a structure that connects the companion to the star (`bridge') and an arc-shaped structure to the west, reminiscent of a spiral arm. 
Spiral arms are now commonly seen in protoplanetary discs and can be related to the presence of planets within those discs \citep[see e.g.][]{boccaletti_possible_2020,  Haffert_2019, ginski2025_spiralarms, Hashimoto_2020}. However, we do not find any clear evidence of a large disc ($\gtrsim$70 au) around V2376 Ori in which the companion V2376 Ori b would be embedded. 
The absence of such a disc is also consistent with the fact that the large mass-ratio of the sources imply a rapid tidal truncation of a large circumstellar disc \citep{Harris_2012_tidal_truncation}.
Thus, it appears more likely that the observed `arc' and bridge are not related to the PMC being embedded in a large or circumbinary protoplanetary disc, but rather to some form of dynamical interaction between the two sources that results in mass transfer from their individual (small) accretion discs. 
Tidal interaction between young objects are shown to indeed create bridges and arcs similar to those observed in Fig.~\ref{fig:muse} \textbf{C} \citep{Clarke_1993, Cabrit_2006_RWAuri}. 
A surprising aspect is that the structure is only seen in certain oxygen lines, not in any of the other atomic lines reported in Table \ref{tab:linelist}. This could reflect particular physical conditions leading to the excitation of these specific lines in the shocked gas or an over-abundance of oxygen with respect to other elements. Additional spectroscopy will be required to understand this.

\section{Conclusion}

Combined observations from JWST and the VLT have allowed us to confirm the presence of a PMC with a mass of $M \simeq \rm 20~M_{Jup}$ around the star V2376 Ori, located in the Orion D association. This result highlights the potential of JWST to identify new PMCs through NIRCam photometric imaging surveys of young associations. Our MUSE observations suggest the presence of a tidal interaction between the circumstellar and circumplanetary discs. Future observations, for instance with SPHERE at VLT, to search for counterparts of the bridge and arc in dust scattering and 
JWST to better characterize the spectral types of the star and companion as well as disc properties are required. The upcoming ELT, with its more than tenfold improvement in angular and spectral resolution in the visible and near-IR, will be crucial to further characterize such dynamical interactions in young planetary systems.

\begin{figure}
    \centering
    \includegraphics[width=\linewidth]{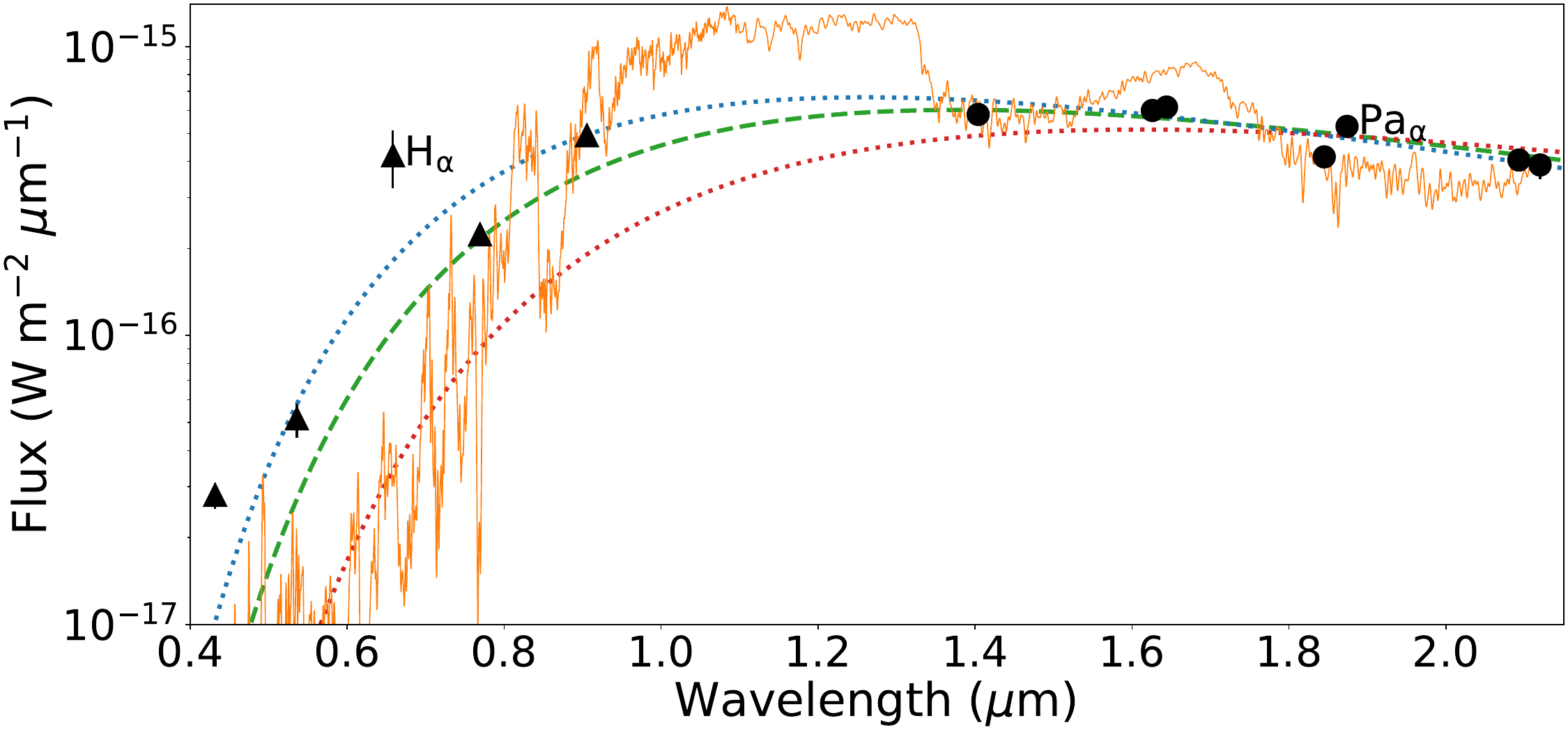}
    \caption{
    Spectral energy distribution (SED) of V2376 Ori b from HST (triangles) and JWST (dots) photometry, shown as black markers with error bars (invisible when smaller than markers). The dashed green line is the best-fit blackbody ($T = 2100$ K), with dotted red and blue lines showing the $T = 1800$ K and 2300 K limits. The orange line shows the best-fit BT-Cond model ($T = 2400$ K, $\log g = 4$).
    }
    \vspace{-0.5cm}
    \label{fig:SED}
\end{figure}

\begin{acknowledgements}
The data were obtained from the Mikulski Archive for Space Telescopes at the Space Telescope Science Institute, which is operated by the Association of Universities for Research in Astronomy, Inc., under NASA contract NAS 5-03127 for JWST. These observations are associated with programs \#1288.
OB, IS are funded by the Centre National d'Etudes Spatiales (CNES) through the APR program. 
This research received funding from the program ANR-22-EXOR-0001 Origins of the Institut National des Sciences de l’Univers, CNRS. 

TJH  acknowledges a Dorothy Hodgkin Fellowship, UKRI guaranteed funding for a Horizon Europe ERC consolidator grant (EP/Y024710/1) and UKRI/STFC grant ST/X000931/1.      

This project is co-funded by the European Union (ERC, SUL4LIFE, grant agreement No101096293). AF thanks project PID2022-137980NB-I00 funded by the Spanish Ministry of Science and Innovation/State Agency of Research MCIN/AEI/10.13039/501100011033 and by "ERDF A way of making Europe".

The authors acknowledge G. D. Marleau and Y. Aoyama for useful discussions on the conversion of H recombination lines intensities into accretion rates.

\end{acknowledgements}

\bibliographystyle{aa} % style aa.bst
\bibliography{aanda.bib} % your references 

\begin{appendix}

\section{Data presentation and reduction}\label{sect:data}

We used the JWST NIRCam images of the ONC from the PDRs4All Early Release Science (ERS) program (\citealt{berne_pdrs4all_2022}, PID 1288) that were previously presented in \citet{habart_pdrs4all_2024}. 
We selected the short-wavelength channel (i.e. for $\lambda\leq2.4\:\mu$m), which includes seven filters (F140M, F162N, F164N, F182M, F187N, F210M, and F212N) with a 0.03'' angular resolution. 
Details on the data reduction procedure to obtain these images are given in \citet{habart_pdrs4all_2024}. The images themselves are available through the specific PDRs4All MAST archive\footnote{\url{https://mast.stsci.edu/portal/Mashup/Clients/Mast/Portal.html}} and through the REGARDS database\footnote{\url{https://regards.osups.universite-paris-saclay.fr/user/jwst}}.
We also use the \textit{Hubble} Space Telescope (HST) images from the ACS/WFC of V2376 Ori 
presented in \citet{ricci_hubble_2008}. This includes five filters (F435W, F555W, F658N, F775W, and F850LP), covering a wavelength range from 0.435 to 0.85 $\mu$m.

The VLT MUSE observations were taken on February 27, 2025 (DDT P114 proposal 114.28JY). The observations originally consisted of four dithers, each with an integration time of 1800 sec and 90 degrees rotation between them. Due to a technical loss only the first dither position was successfully observed, yet, after inspection, the quality of this single exposure was deemed sufficient for this initial analysis.

The data were reduced using the ESO MUSE data reduction pipeline (v.2.10.14) together with the python wrapper, which is part of MUSEpack \citep{Zeidler2019, Zeidler2019a}. We apply all steps of the MUSE data reduction pipeline using the ESO provided master calibrations (\texttt{MASTERBIAS}, \texttt{MASTERDARK}, \texttt{MASTERFLAT}). To subtract the background we use the standard method `model' using 10\% of the datacube spaxels (\texttt{fraction=10}) after ignoring the darkest 10\% (\texttt{ignore=10}). We visually inspect the skymask to ensure that it selects the darkest regions of our cube while rejecting all artifacts. Furthermore, we carefully inspect the sky spectrum and sky continuum to ensure that only telluric lines are removed from the final datacube. Once all data will be available we will combine these with the current dataset, which will further enhance our data product by allowing us to remove any remaining artifacts and significantly increase the signal-to-noise.

\section{Position of the sources}

\subsection{Relative position between HST and JWST images}\label{sect:rel_position}
\begin{figure}
    \centering
    \includegraphics[width=\linewidth]{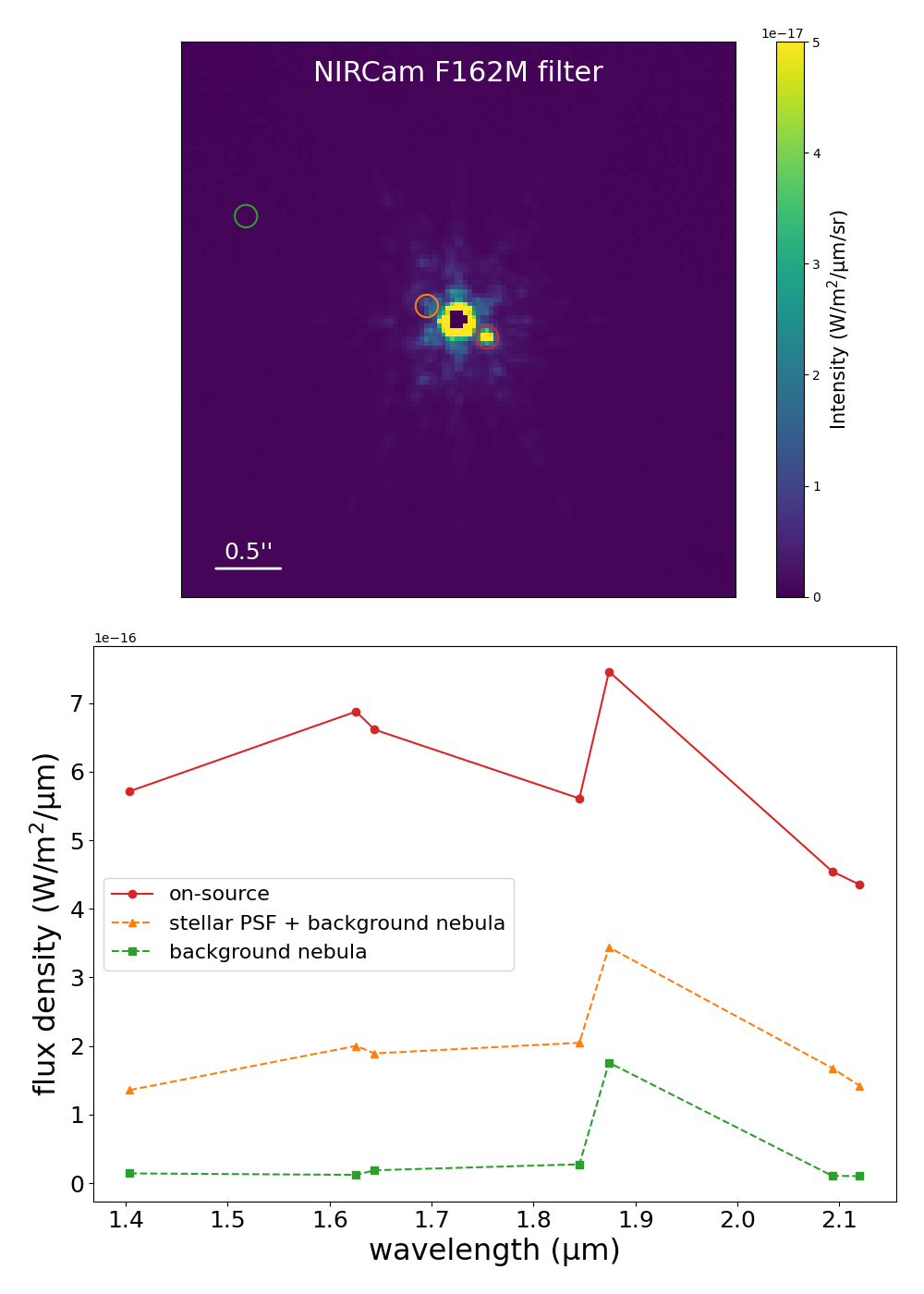}
    \caption{{
     Top: NIRCam image in the F164N filter. The red circle corresponds to the aperture used to extract the companion's SED (on-source SED). The orange circle, placed at the opposite of  the companion with respect to the star, represent the apertures used to estimate the SED of the stellar PSF + background nebula. The green circle corresponds to the aperture of the background nebula SED. The radius of these apertures is 0.08''.
    Bottom: In red, On-source SED. In orange, SED of the background nebula + stellar PSF. In green, the SED of the background nebula emission.    
    } }
    \label{fig:SED_comparison}
\end{figure}

 {
The proper motion of V2376 Ori provided by Gaia DR3 is $\delta$ra=-0.148 mas/yr, $\delta$dec=-1.082mas/yr. Even considering the largest temporal span of nearly 21 years between HST and VLT, the proper motion is expected to be smaller than one pixel, making common proper motion confirmation difficult. Therefore we only analyse the relative position of V2376 Ori and the companion at two different epochs using the HST F435W image (taken in October 2004) and JWST F187N  (taken in September 2022) images.
We exclude the MUSE image from this analysis because of the lack of additional sources in the field of view preventing robust astrometric alignment.
Figure~\ref{fig:rel_position} shows the contours of the HST F435W  (blue continuous lines) and JWST F187N (dashed red lines) images. Both are centered on V2376 Ori, and oversampled on a common grid to ensure alignment to about a tenth of a NIRCam pixel (0.003'').
The companion (to the southwest of V2376 Ori) is shifted by $\rm \Delta ra$ = 9 mas, $\Delta$dec = 25 mas between the two epochs, i.e. over nearly 18 years.
Considering that the pixel size is 50 mas for \textit{Hubble} and 30 mas for NIRCam, and that the source shows a non-symmetric shape in the \textit{Hubble} image, this difference is compatible with the absence of relative movement of the two sources over this time period.

\begin{figure}[t]
    \centering
    \includegraphics[width=0.85\linewidth]{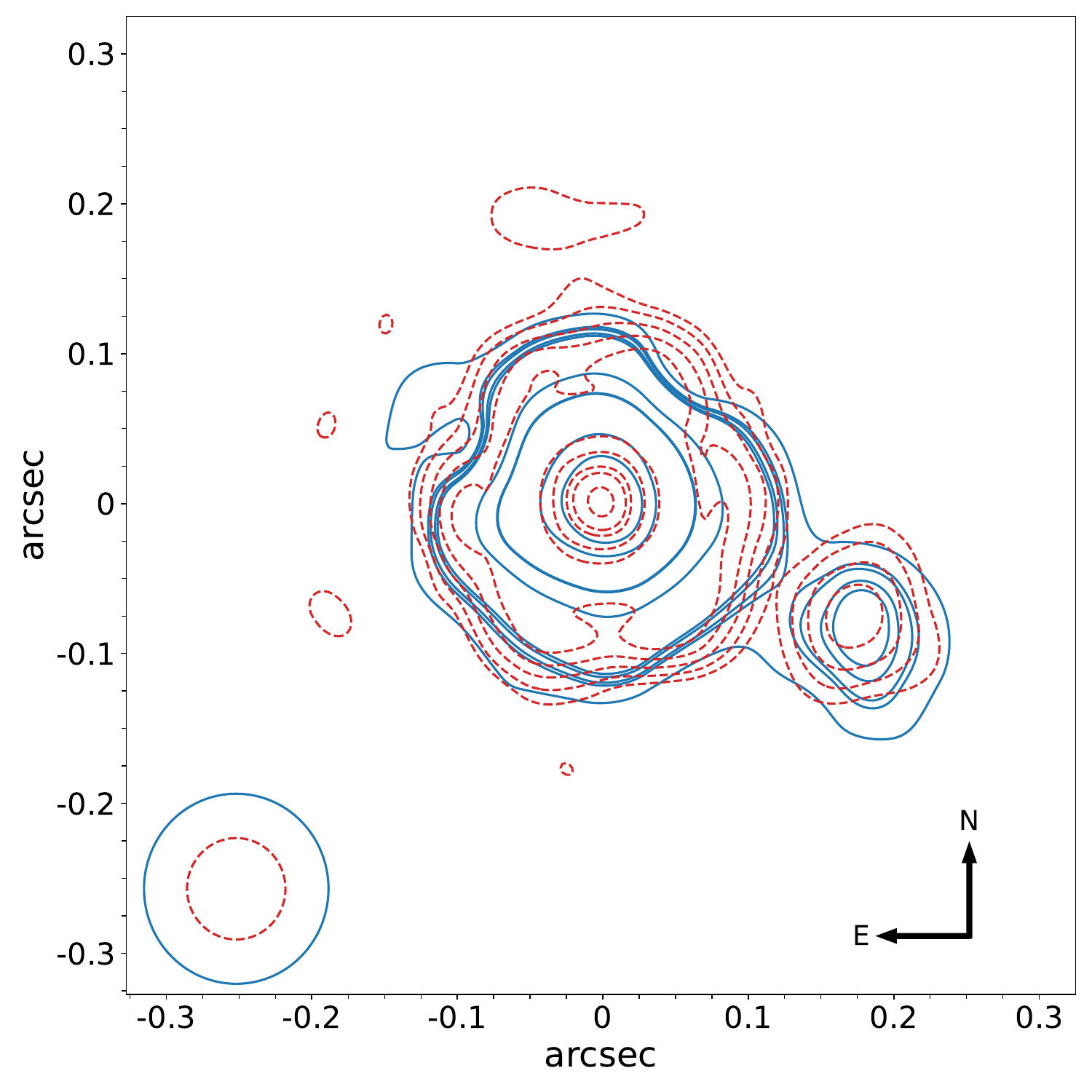}
    \caption{Contours of the HST F435W image (continuous blue lines) and JWST F187N image (dashed red lines). Both images are north-aligned, centered on V2376 Ori and oversampled on a grid with pixel sizes of 0.003''. The full width half maxima of the point spread functions are shown with circles in the bottom-left corner, for each instrument.}
    \label{fig:rel_position}
\end{figure}

\subsection{Angular separation of the sources}\label{sect:ang_separation}

We estimated the angular separation between V2376 Ori and its companion in the HST F435W image, JWST F164N image and in the integrated MUSE cube.
We first extracted an intensity profile along a cut performed along the axis connecting the two sources. 
We then determine the position of V2376 Ori and its companion in the profile with the highest and second highest intensity peak, respectively.
The distance in pixels between these two peaks is finally converted into an angular separation considering the respective pixel sizes in arcsec of the different instruments\footnote{the pixel sizes for ACS/WFC HST, NIRCam JWST and VLT MUSE data are respectively 0.05'', 0.031'', and 0.025''}.
This yields for the F435W HST image, F164N JWST image and the MUSE image an angular separation of 0.23$\pm$0.05, 0.25$\pm$0.03'' and 0.23$\pm$0.03'', respectively.
The final angular separation $a$ is considered as the mean of these values, leading to $a$ = 0.23$\pm$0.03''.

\section{Extraction of the spectral energy distribution}\label{sect:PSF_sub}

Figure \ref{fig:SED} shows the SED of V2376 Ori b extracted from HST-ACS/WFC and JWST-NIRCam images. 
The flux arising from V2376 Ori b is contaminated by the background nebula emission from the ONC (see Fig.~\ref{fig:kaleidoscope} {\bf C}) and by the diffraction pattern (PSF) due to V2376 Ori (see, e.g. Fig.~\ref{fig:kaleidoscope} {\bf F}, {\bf G}).
The nebular emission is estimated using the \texttt{Background2D} function from the \texttt{Photutils} Python library\footnote{https://photutils.readthedocs.io/en/2.1.0/user\_guide/index.htm}.
We use 50$\times$50 pixel boxes to create a low-resolution map with a sigma-clipping of three, which is then filtered with a 2D median filter of 5$\times$5 pixels and interpolated to obtain the final 2D background. The images are then subtracted from this 2D background.
Next we use three different approaches to remove the contribution of the diffraction spikes: 
\begin{enumerate}
    \item For each image, we subtract a version rotated by 180° around the stellar center, so that every diffraction spike is cancelled by its counterpart on the opposite side of the star.  
    \item We subtract each image by its symmetric counterpart along the X-detector axis, and along the Y-detector axis. We then average the results obtained along both axes.
    \item We use the \texttt{STPSF} python package\footnote{https://stpsf.readthedocs.io/en/latest/} to create a stellar PSF model for each JWST-NIRCam filter and then subtract it directly from the corresponding image. This method is only used for the JWST NIRCam dataset. 
\end{enumerate}
Finally, the on-source flux in each background-subtracted and PSF corrected image is extracted using the \texttt{AperturePhotometry} function from the \texttt{Photutils} library with an aperture radius of 0.08'' centered on V2376 Ori b, which is shown in Fig.~\ref{fig:kaleidoscope} ({\bf G}) with the dashed red circle. 
The final flux in a given filter is the mean of the fluxes obtained for the different PSF subtraction methods.
Figure~\ref{fig:SED_comparison} presents the observed emissions at the position of the companion, at a position symmetric to that of the companion wrt to the star representing the contribution from the PSF (+background) at the companion position, and the emission from background nebula. 
The background nebula contributes to the on-source flux at a level of less than $5~\%$ in most filters (except the F187N filter at 1.87~µm where this contribution is $\sim~24~\%$). The stellar PSF contributes to at most $\sim~30~\%$ of the on-source flux. This shows that the on-source flux is mainly affected by the stellar PSF rather than by the background nebular emission, but still largely dominated by emission from V2376 Ori b itself. Since the PSF and background contributions are both smaller than the flux of V2376 Ori b, we expect that residual emission after subtraction contributes to at most a few $\%$.}

\section{Physical properties of the V2376 Ori system}\label{properties}

\begin{figure}[h]
    \centering
    \includegraphics[width=0.9\linewidth]{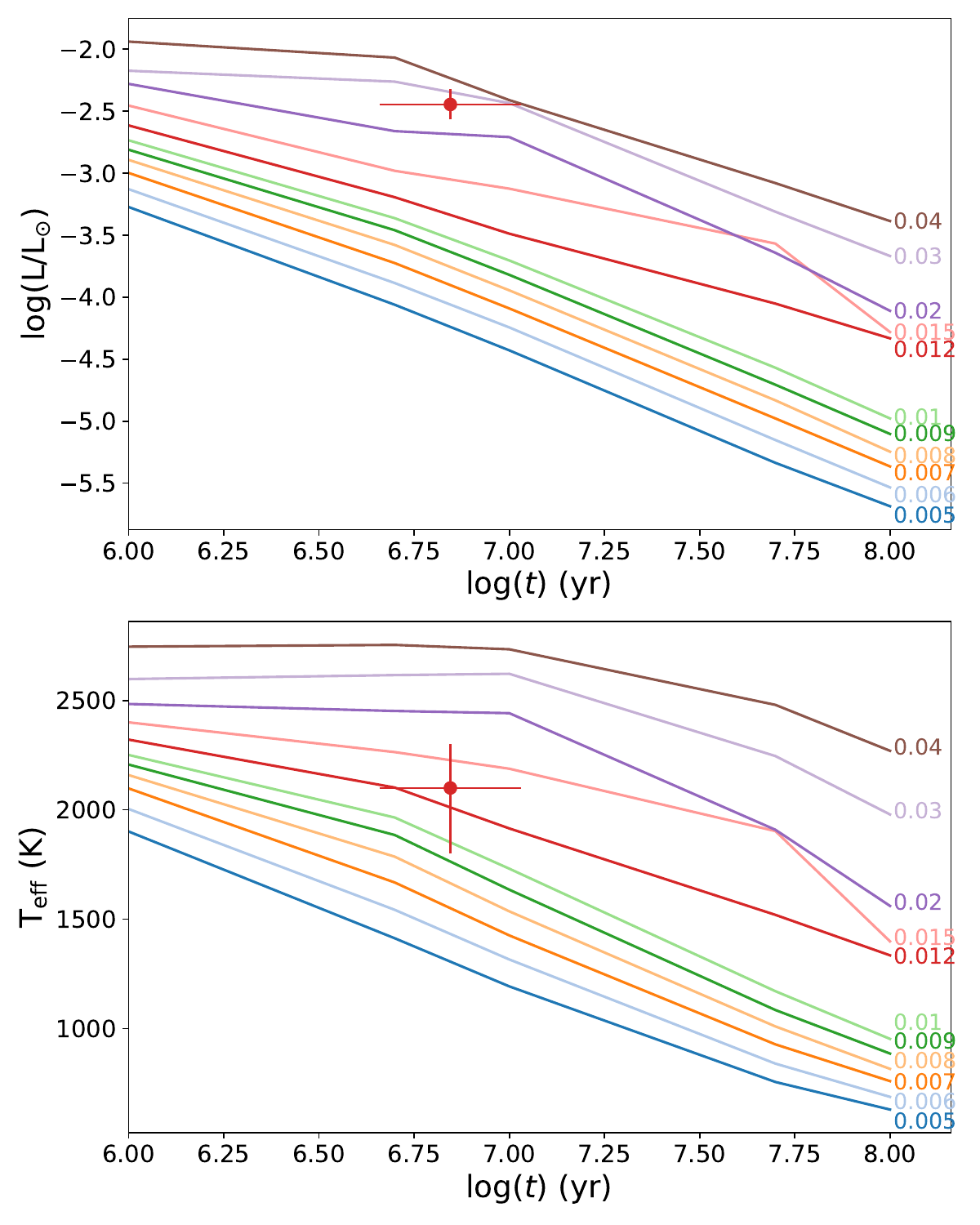}
    \caption{
    Evolutionary tracks from the model of \citet{baraffe_evolutionary_2003} showing the temporal evolution of (top) the luminosity and (bottom) the effective temperature ($T_{\rm eff}$) for cool brown dwarfs and giant exoplanets. Each curve corresponds to a specific mass (in $\rm M_\odot$), as labelled at the end of the tracks. The red data point represents the companion V2376 Ori b, plotted with its age and luminosity (top panel), and its effective temperature (bottom panel), with associated uncertainties, using the black body model fitted to the SED. 
    }
    \label{fig:evolutionary_tracks}
\end{figure}

Table~\ref{tab:Companion_prop} summarizes the physical properties of the companion V2376 Ori b.
Figure~\ref{fig:evolutionary_tracks} shows the luminosity (top panel) and temperature (bottom panel) of V2376 Ori b, used to estimate its mass according to evolutionary tracks from \citet{baraffe_evolutionary_2003}. 
The derivation of the companion's luminosity is described in Sect.~\ref{sect:temperature_lumin_mass}. 

\begin{figure} [h]
    \centering
    \includegraphics[width=\linewidth]{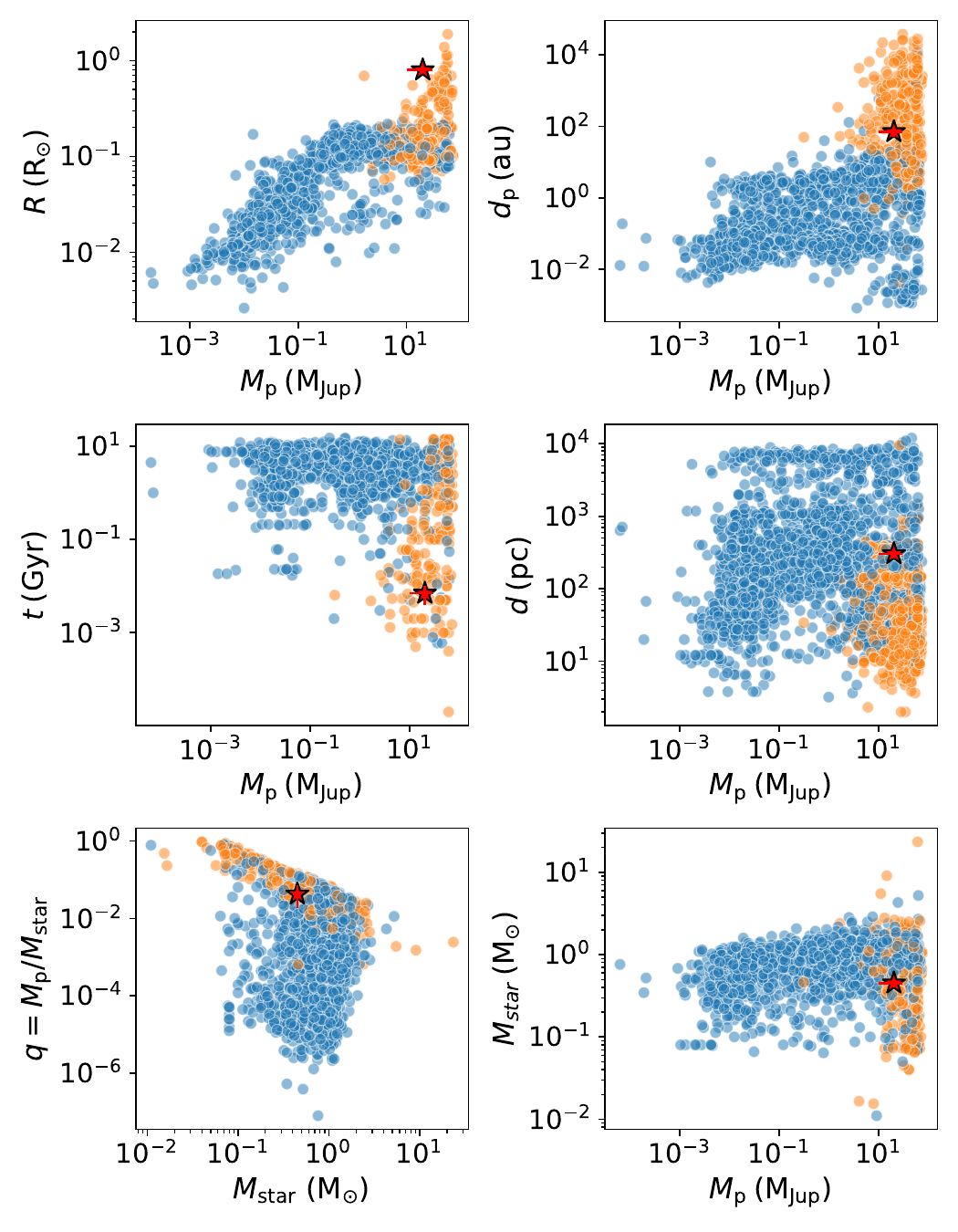}
    \caption{Comparative placement of V2376 Ori b with known PMCs and exoplanets based on several properties. V2376 Ori b is shown as a red star-shaped marker, associated error bars are included but may be smaller than the marker size and thus not visible. Blue points represent exoplanets detected via indirect methods (e.g. radial velocity or transits), while orange points correspond to companions identified through direct imaging. All comparative data is sourced from the portal \texttt{exoplanet.eu} of The Extrasolar
    Planets Encyclopaedia.}
    \label{fig:exoplanets}
\end{figure}

We estimate the companion's mass to be in the range $\rm 10-30\:M_{Jup}$, and the mass of the primary to be $\rm 0.4-0.5\:M_{\odot}$. These values correspond to a companion-to-stellar mass ratio $q = 0.02-0.075$. 
In Fig.~\ref{fig:exoplanets}, we compare all the derived physical properties of the V2376 Ori b system to those of known exoplanets from the portal \texttt{exoplanet.eu}\footnote{\url{https://exoplanet.eu/}} of The Extrasolar Planets Encyclopaedia.


\begin{table}[h]
    \caption{Properties of V2376 Ori b.}
\centering
    \begin{tabular}{ccc}
    \hline
    Parameter (unit)   &  Symbol  &  Value \\ \hline\hline
    Distance  (pc)  & $d$      &    304$\pm$42 \\
 %   Angular separation ('')  & $a$  &    0.23$\pm$0.03 \\
    Projected separation (au)  &  $a$  &    71$\pm$21 \\
    Age (Myr)  & $t$        &   $7\pm3$ \\ 
    Mass (${\rm M_{ Jup}}$)      &     $M_{\rm p}$          &  10-30  \\ 
    Radius ($\rm R_{\odot}$)     &  $R_{\rm p} $
    & ${ \rm0.8\pm0.1}$ \\ 
    mass ratio (companion/star) & $q$ & 0.02-0.075 \\
    Effective temperature (K) & $T_{\rm eff}$  & $2100^{+200}_{-300}$ \\
    Luminosity ($\log(\text{L}_{\odot})$) & $\log(L)$   & $-2.4\pm0.1$ \\
    $\rm H\alpha$-luminosity ($\log (\text{L}_{\odot})$)  &   $ \log(L_{\rm H\alpha})$ & $-5.6\pm0.2$    \\
    Accretion luminosity ($\log( \text{L}_{\odot})$)   &  $ \log(L_{\rm acc})$   &  $-3.7\pm0.2$ \\
    Accretion rate ($\rm \log (M_{Jup}\:yr^{-1})$) &  $ \log(\dot{M}_{\rm acc})$ &  ${\rm -6.6\pm0.2}$\\
    
    \hline
    
    \end{tabular}

    \label{tab:Companion_prop}
\end{table}

\section{MUSE spectra}\label{sect:linelist}

\subsection{Extraction of MUSE spectra}

\begin{figure}[h]
    \centering
    \includegraphics[width=0.8\linewidth]{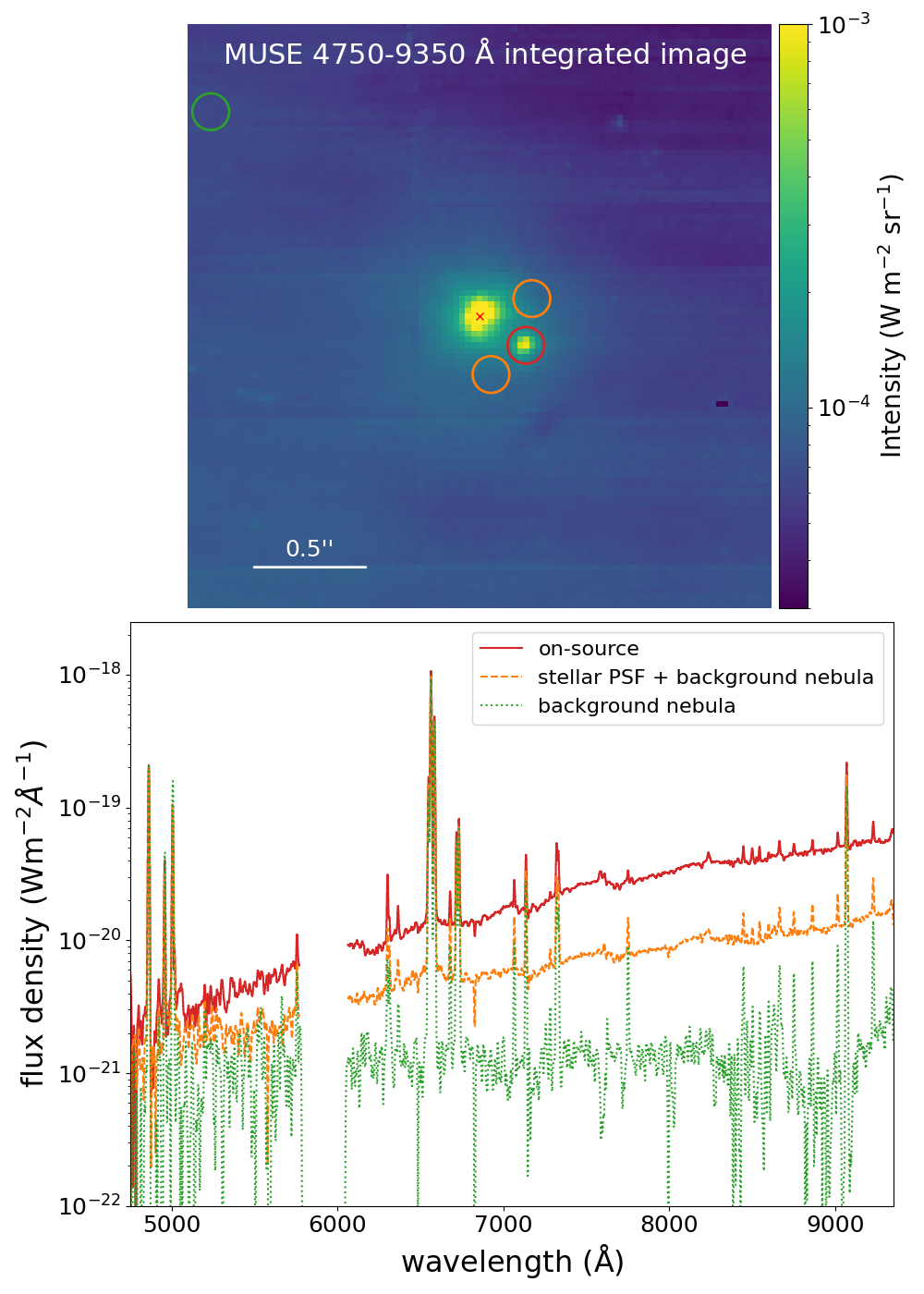}
        \caption{{Top : MUSE image integrated over the range 4750-9350 $\rm \AA$. The red circle corresponds to the aperture used to extract the companion's spectrum (on-source spectrum). The orange circles, placed at equi-distance of the star and the companion, represent the apertures used to estimate the spectrum of the stellar PSF (+ background). The green circle corresponds to the apertures of the background nebula spectrum. The radius of these apertures is 0.08''. Bottom : In red, the on-source spectrum. In orange, the mean of the stellar PSF (+background) spectra. In green, the background nebula spectrum.}}
    \label{fig:bkg_sub_spectrum}
\end{figure}
{
The spectrum of V2376 Ori and its companion are extracted from the MUSE cube. For the star, we use a circular aperture of 0.1''  radius centered on the source. We then subtract the background nebula emission, estimated within two circular apertures of 0.1''  radius, placed on either side of the star, in the direction perpendicular to the star-companion axis.
For the companion, we perform the subtraction using two apertures situated on the sides of V2376 Ori b following an approach similar to \citet{eriksson_strong_2020} for Delorme 1 (AB)b.  
The used apertures are shown on Fig.~\ref{fig:bkg_sub_spectrum}. 
The spectrum of V2376 Ori b is obtained by subtracting the on-source spectrum from the average background+stellar PSF contribution. Between 7950 and 8050 $\rm \AA$, the background nebula contributes to the on-source spectrum to $\sim 4$\%, whereas the PSF contributes to $\sim 30\%$. Since the PSF and background contribution are both smaller than the flux of V2376 Ori b, we expect that residual emission after subtraction contributes to at most a few $\%$ in the continuum.
The mean of the two obtained background+PSF spectra is then subtracted from the on-source spectrum to obtain the final spectrum of the companion.}
The direct subtraction leads to artifacts with negative intensities and P-cygni-like profiles, possibly produced by velocity effects and/or thermal broadening. We apply a gaussian filter with a standard deviation of 2.5 pixels to remove these effects.
The obtained spectra are then calibrated using the photometric fluxes derived in the same aperture in the F775W HST filter using the \texttt{MPDAF} Python package and associated filter throughput. This filter's wavelength range is fully covered by MUSE and is not dominated by the strong $\rm H\alpha$ emission line.

\subsection{Emission lines in V2376 Ori b spectrum}

We extract the flux of each emission line in the V2376 Ori b spectrum by fitting Gaussian profiles using \texttt{MPDAF} \citep{piqueras_mpdaf_2017}. Table \ref{tab:linelist} summarizes the detected emission lines and their computed fluxes.

\begin{table}[h]
\caption{Identified emission lines in the V2376 Ori b spectrum and their associated Gaussian-integrated fluxes. }
 \begin{threeparttable}
    \centering

        \begin{tabular}{ccc}
    \hline
    Species & Wavelength (\AA) & $F_{line}$ ($10^{-20} \, \text{W/m}^2$) \\ \hline\hline
    H$\beta$\tnote{†}      &   4862 & 7.4 \\ 
    $[\text{O\:{\scriptsize I}}]$ & 6300 & 9.4 \\
    $[\text{O\:{\scriptsize I}}]$ & 6363 & 3.2 \\
    $[\text{N\:{\scriptsize II}}]$ & 6548 & 11\tnote{*} \\
    H$\alpha$ & 6565 & 93 \\
    $[\text{N\:{\scriptsize II}}]$ & 6584 & 18 \\
    HeI & 6678 & 1.4\tnote{*} \\
    HeI & 7065 & 1.9 \\
    $[\text{O\:{\scriptsize II}}]$ & 7320 & 4.6\tnote{*}\\
    $[\text{O\:{\scriptsize II}}]$ & 7330 & 2.0 \\
    HeI & 8235 & 4.8\tnote{*}\\
    OI & 8446 & 4.5 \\
    HeI & 8499 & 7.1 \\
    HI & 8543 & 6.1 \\
    HI & 8599 & 1.1\tnote{*} \\
    CaII & 8662 & 5.9 \\
    $[\text{C\:{\scriptsize I}}]$ & 8727 & 0.8 \\  
    HI & 8751 & 4.2\tnote{*} \\
    HI & 8864 & 2.5\tnote{*} \\
    HI & 9016 & 2.6 \\
    HI & 9230 & 8.4

    \\ \hline
    
    \end{tabular}
    
\tablefoot{Unless otherwise indicated, fluxes are measured on the convolved spectrum.}
     \begin{tablenotes}
        \item [†] this emission may be residual emission from the nebula, as there is significant noise in the spectrum at this wavelength.
       \item [*] non detection in the convolved spectrum: the flux is measured before convolution. 
     \end{tablenotes}

    \label{tab:linelist}
\end{threeparttable}

\end{table}

We detect the [\ion{N}{ii}]~$\lambda$6584, [\ion{O}{i}]~$\lambda$6300,~$\lambda$6363 and [\ion{O}{ii}]~$\lambda$7320,~$\lambda$7330 lines, which are commonly associated with jets and outflows \citep{solf_1993_jets, Hartigan_1995_accretion, fFlores-Rivera_2023}, as well as with externally irradiated discs \citep{haworth_vlt_2023, aru_kaleidoscope_2024}. Notably, we also detect the forbidden [\ion{C}{i}]~$\lambda$8727 line--to our knowledge, the first such detection in a PMC. This line has recently been observed in irradiated discs in Orion \citep{haworth_vlt_2023, aru_kaleidoscope_2024, Aru_2024_CI} and its excitation is attributed to far-UV (FUV) irradiation of neutral gas, either as a result of C+ recombinations \citep{Escalante_1991} or FUV radiative pumping of neutral atomic carbon \citep{Goicoechea_2024}. 
Detection of the Li $\lambda6708$ absorption line is compatible with V2376 Ori b being a young, low-mass object \citep{Magazzu_lithium, Phan_Bao_lithium}. 
The presence of CaII (8662 $\AA$), HeI (6678, 7066 $\AA$), and OI (8447 $\AA$) lines points to a strongly accreting system \citep{white_very_2003, jayawardhana_evidence_2003, Bowler_2014}. 
This is also in line with the strong emission in H$\alpha$, and HeI \citep{jayawardhana_evidence_2003, white_very_2003, eriksson_strong_2020}. 
These accretion tracers have previously been detected in PMCs \citep{muzerolle_detection_2000,  natta_accretion_2004, joergens_ots44_2013, eriksson_strong_2020}.

\section{Spectral type of the host star}\label{sect:host_star}

\begin{figure}[h]
    \centering
\includegraphics[width=0.9\linewidth]{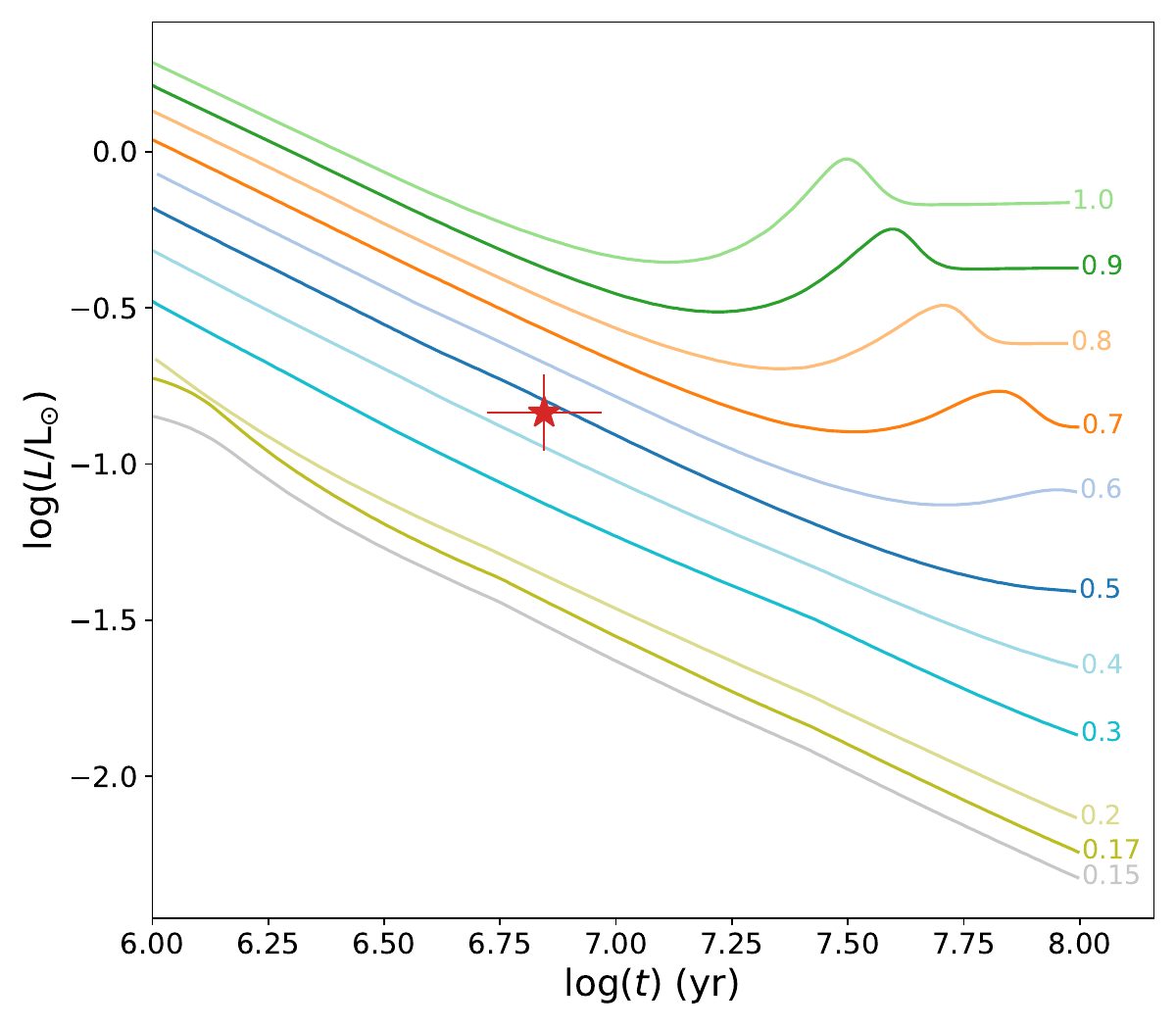}
    \caption{Evolutionary tracks from the model of \citet{baraffe_2015_evolution} showing the temporal evolution of the luminosity for low-mass stars. Each curve corresponds to a specific mass (in $\rm M_\odot$), as labelled at the end of the tracks. The red star marker represents the primary V2376 Ori, plotted with its age and luminosity, with associated uncertainties. Its luminosity is computed from the SED.
    }
    \label{fig:star_tracks}
\end{figure}

\begin{figure}[h]

    \centering
\includegraphics[width=0.95\linewidth]{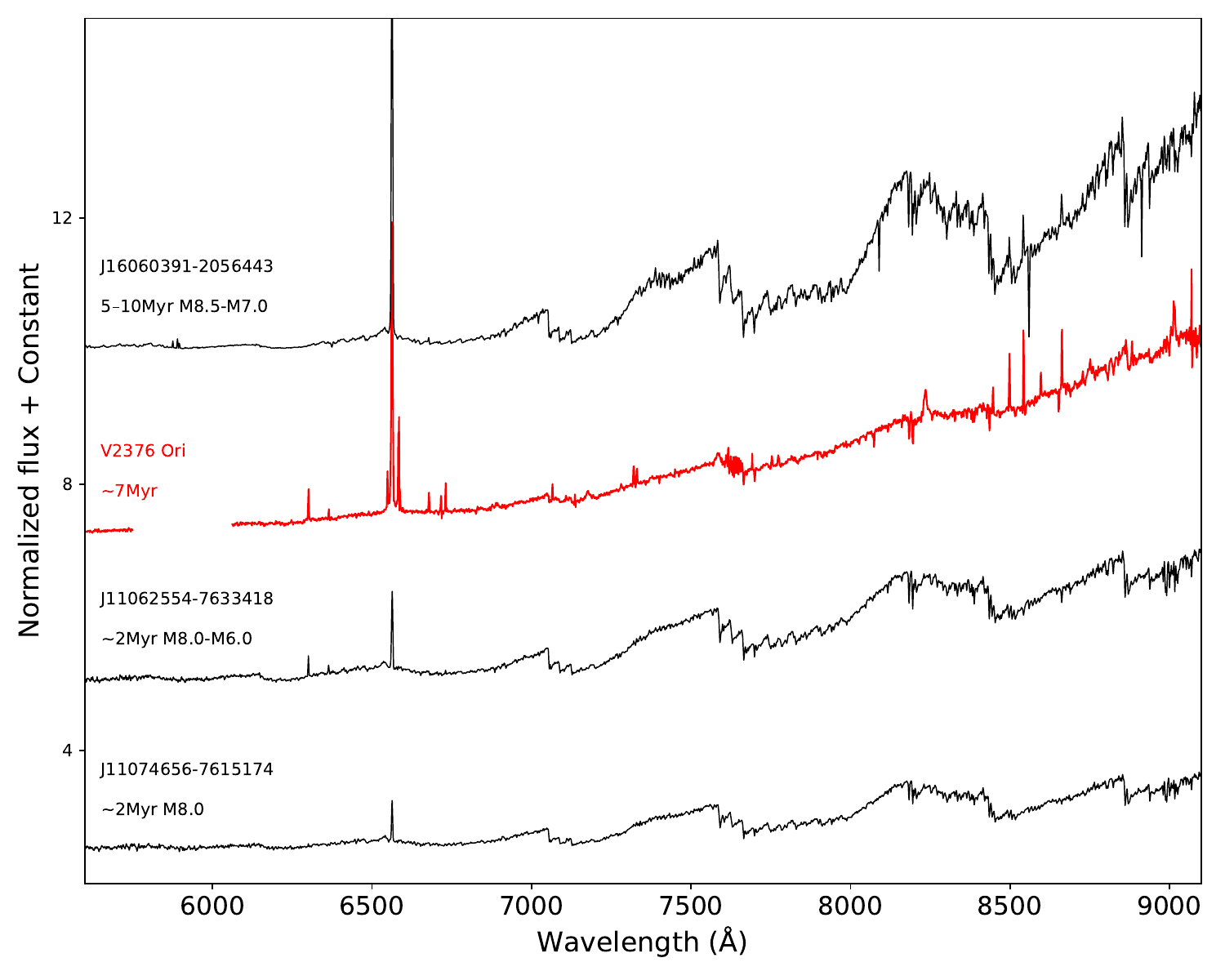}
    \caption{Spectrum of the host star V2376 Ori (solid red line) compared with several young stellar spectra (solid black lines) from \citet{manjavacas_spectral_2020}.}
    \label{fig:manjavacas}
\end{figure}

We tentatively fit the spectrum and the SED of V2376 Ori with the \texttt{Phoenix} model \citep{husser_phoenix_2013} and with a simple black-body. The best-fits obtained are very poor, with a reduced $\chi^2$ of $\approx 330$ and $\approx 20$, respectively.
In Fig.~\ref{fig:manjavacas}, we present a comparison of the MUSE spectrum of V2376 Ori to stellar templates from low mass, young, M-type stars presented in \citet{manjavacas_spectral_2020}.
In terms of slope, the V2376 Ori spectrum is similar to an M8-M6.0 spectral type, however with rather weak molecular absorption bands, which is more compatible with the spectra of the youngest objects in Fig.~\ref{fig:manjavacas}. 
These spectral types suggest a mass of $\sim$0.3 M$_{\odot}$, smaller than what has been derived from the luminosity and evolutionary tracks (Fig. \ref{fig:star_tracks}), that is, $M\simeq 0.5$M$_{\odot}$. Spectral type and mass/luminosity can be reconciled if the system is a binary that is unresolved with MUSE.

\section{Accretion rate}\label{sect:accretion}

The presence of several accretion tracers and the young age of the system ($7\pm3$ Myr) strongly suggest that accretion is ongoing in V2376 Ori b. 
We measure the $\rm H\alpha$ flux by performing a Gaussian fit on the convolved spectrum using the tools provided in the MPDAF package developed by \citet{piqueras_mpdaf_2017}. The measurement is based on the degraded-resolution spectrum shown in Fig.~\ref{fig:muse}D, and yields a flux of $F_{\rm H\alpha} = 92.8 \pm 0.2 \times 10^{-19}\:\rm W/m^{2}$. 
Considering this H$\alpha$ flux with a distance to the source of 304 pc, this leads to $L_{\rm H\alpha} = 10^{-5.6\pm0.2}\:\rm L_{\odot}$.
To convert this $\rm H\alpha$ luminosity into an accretion luminosity, we adopt the planetary surface shock model from \citet{aoyama_comparison_2021} that considers the contribution of shocks between the circumstellar disc and the circumplanetary disc to the total H$\alpha$ emission:
\begin{equation}\label{eq:Aoyama2021}
    \log \left( \dfrac{L_{\rm acc}}{\rm L_\odot} \right) = (0.95\pm0.006)\times \log \left(\dfrac{L_{\rm H\alpha}}{\rm L_\odot}\right) + (1.61\pm0.04).
\end{equation}
This relation yields an accretion luminosity $L_{\rm acc} = 10^{-3.7\pm0.2}\:\rm L_{\odot}$. Considering a radius $R_{\rm p}=0.8\: \rm R_{\odot}$ and a mass $M_{\rm p} = 20 \:\rm M_{Jup}$ (Table \ref{tab:Companion_prop}) this leads to a mass accretion rate $\dot{M}_{\rm acc}=10^{-6.6\pm0.2}\:\rm M_{Jup}/yr$.

\section{Atmospheric model fitting to V2376 Ori b MUSE spectrum}\label{sect:models}

\begin{figure}[!h]
    \centering
    \includegraphics[width=\linewidth]{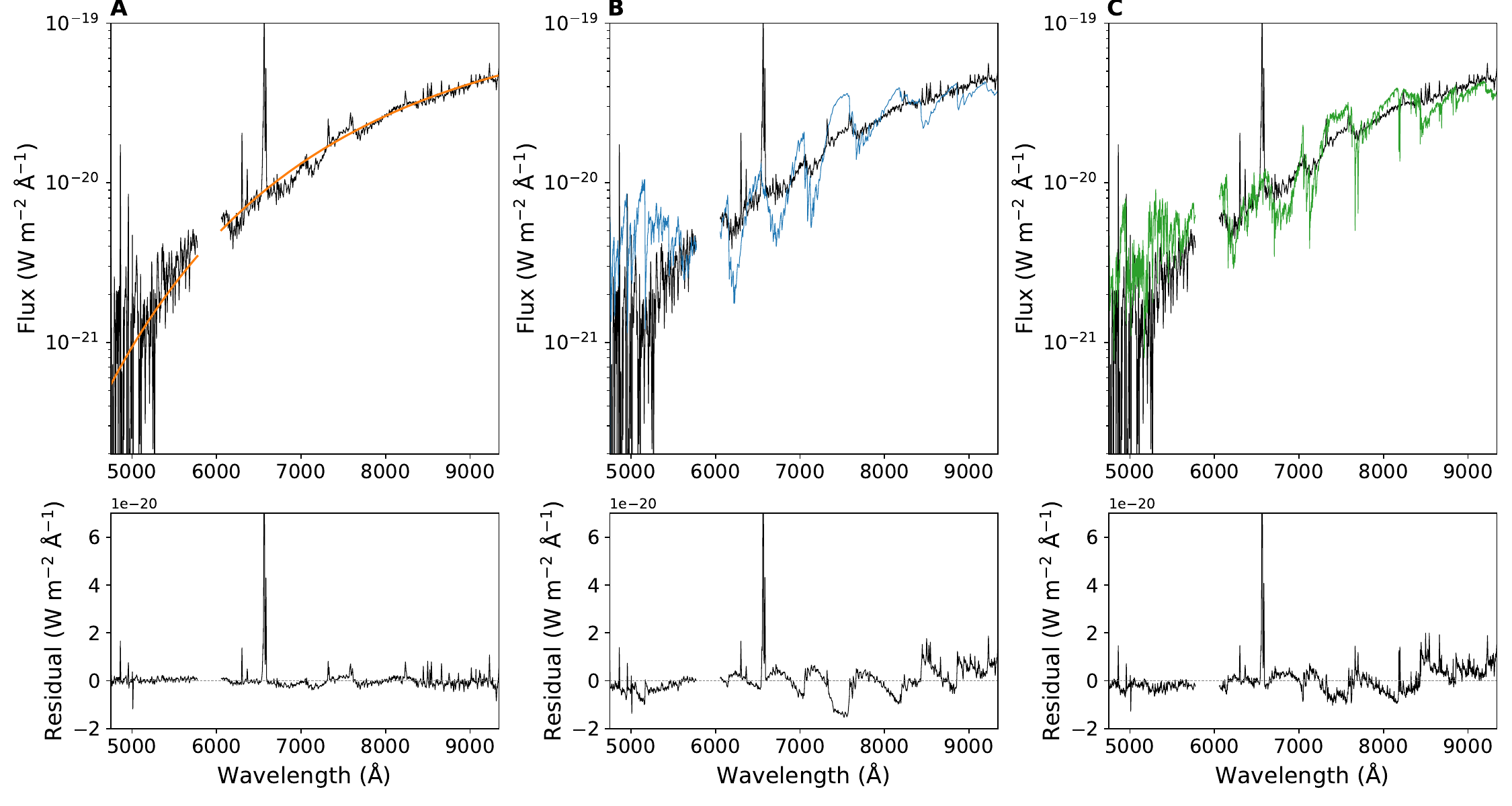}
    \caption{Spectrum of V2376 Ori b (black curve) compared to three models, with the corresponding residuals shown below each fit. (\textbf{A}) orange line shows a blackbody fit at $\rm T = 1900\pm300\:K$.  (\textbf{B}) comparison with the ATMO 2020 synthetic spectrum. The blue line shows the best-fit for $\rm T = 3000\:K$ and $\log(g)=4.0$. (\textbf{C}) comparison with the BT-Cond model. The best-fit for $\rm T=3000\:K$ and $ \log(g)=5.0$ is represented by the green line. Residuals are computed as the difference between the observed spectrum and each model.
    }
    \label{fig:residus}
\end{figure}

\begin{figure}
    \centering
    \includegraphics[width=0.8\linewidth]{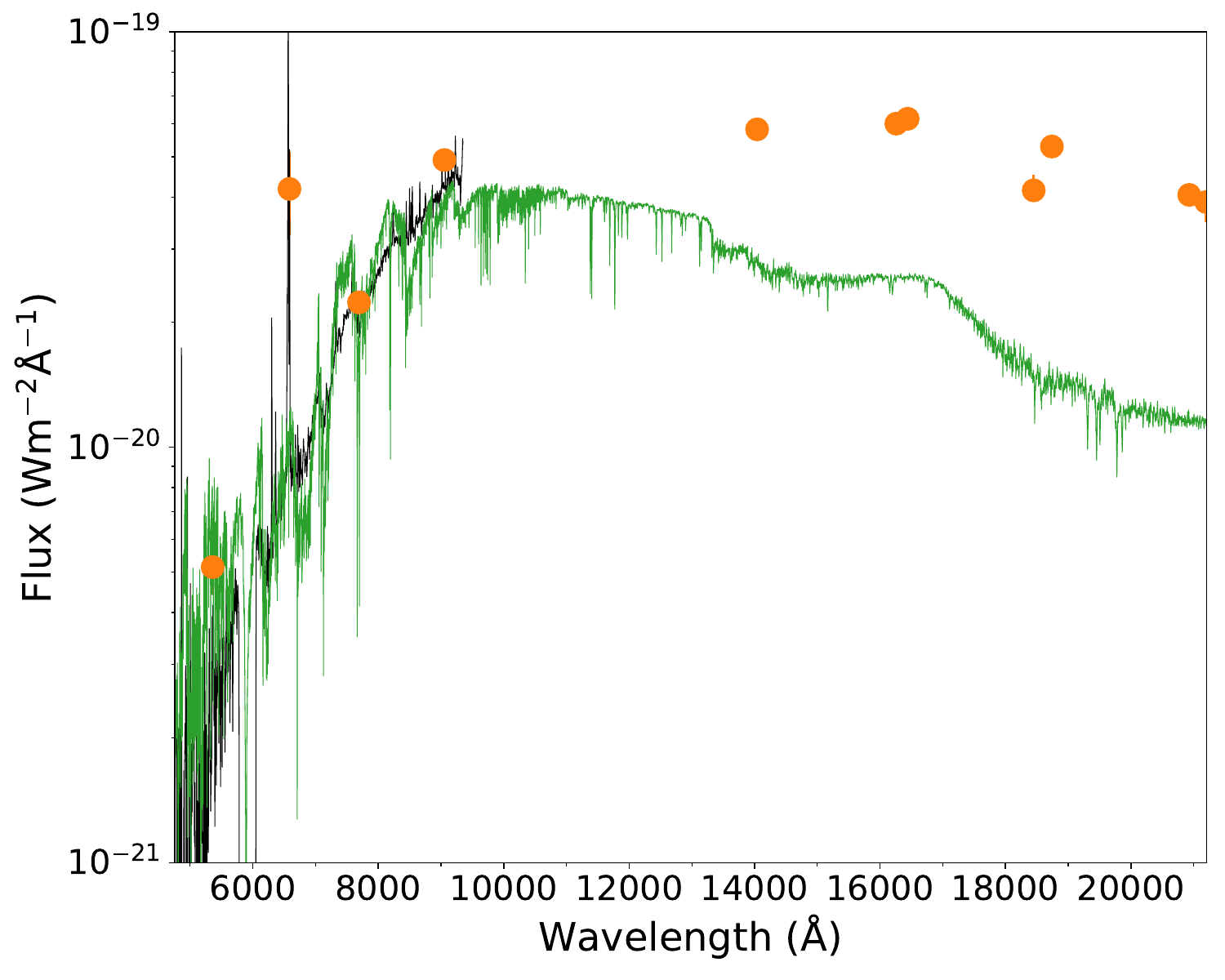}
    \caption{Spectrum (black curve) and SED of V2376 Ori b (orange data points) compared with the best-fit of the spectrum using the BT-Cond model (green curve). The best-fit for $\rm T=3000\:K$ and $ \log(g)=5.0$.
    }
    \label{fig:BTCond_spec}
\end{figure}

Figure~\ref{fig:residus} presents the best-fits models obtained for the V2376 Ori b MUSE spectrum using a blackbody emission, 
the ATMO2020 model \citep{phillips_new_2020}, and the BT-Cond model \citep{Allard_2001, Allard_BTCond_2011, Allard_2012}.
Among these three models, the black-body (panel~\ref{fig:residus}{\bf A}) provides the best fit (and fewer number of free parameters) with a photosphere effective temperature of 1850 K.
The ATMO2020 model (panel~\ref{fig:residus}{\bf B}) hardly reproduces the observed spectrum. Moreover, the best-fit yields an effective temperature of 2900 K for the source, which is difficult to reconcile with the observed luminosity and out of the range of the recommended temperatures for ATMO. 
The BT-Cond models (panel~\ref{fig:residus}{\bf C}), which are more suited for high-temperature, low-mass objects, better reproduce the spectrum than ATMO2020, but still present higher residual amplitudes than a simple blackbody. The effective temperature of 2900 K derived from this model appears significantly higher than the previous estimate made from the SED in Sect. \ref{sect:temperature_lumin_mass} and implies an underestimated flux in the JWST filters as shown in Fig.~\ref{fig:BTCond_spec}. 
Overall, this comparison suggests that current atmospheric models are still not fully adapted to young, low-mass objects such as V2376 Ori b. This can be attributed to the limited number of such objects observed and thoroughly characterized to date.

\end{appendix}

\end{document}